\documentclass[12pt]{iopart}
\usepackage{amsfonts}
\usepackage{amssymb}
\usepackage[dvips]{graphicx}
\eqnobysec
\newcommand{\dd}{\delta}

\newcommand{\be}{\begin{equation}}
\newcommand{\ee}{\end{equation}}
\newcommand{\bea}{\begin{eqnarray}}
\newcommand{\eea}{\end{eqnarray}}
\newcommand{\nn}{\nonumber}

\newcommand{\bse}{\begin{subequations}}
\newcommand{\ese}{\end{subequations}}
\begin{document}
\title{Quantum discrete Dubrovin equations}
\author{Chris M Field and Frank W Nijhoff}
\address{Department of Applied Mathematics, University of Leeds, Leeds LS2 9JT, UK}
\eads{\mailto{cfield@maths.leeds.ac.uk} and \mailto{frank@maths.leeds.ac.uk}}
\begin{abstract}

\noindent
The discrete equations of motion for the quantum mappings of KdV type are given in terms of the
Sklyanin variables (which are also known as quantum separated variables). Both temporal
(discrete-time) evolutions and spatial (along the lattice at a constant time-level) evolutions
are considered. In the classical limit, the temporal equations reduce to the (classical)
discrete Dubrovin equations as given in a previous publication (Nijhoff F W 2000
 \textit{Chaos, Solitons and Fractals} \textbf{11} 19-28). 
The reconstruction of the original dynamical variables in terms of the Sklyanin variables is also
achieved.

\end{abstract}


	\section{Introduction}\label{qddintro}

The quantization of discrete-time systems is an outstanding problem within the wider context of the 
quantum theory of dynamical systems.  From the 
perspective of quantum chaos theory (quantum chaology), quantum mappings have been studied by Berry \textit{et al}. \cite{ber:quantmap}.
Quantum mechanical systems evolving in discrete-time have also been considered by
Bender \textit{et al}. \cite{Bender:Discretesurv}, \cite{Bender:DiscreteQM}, where the emphasis was on 
the application of finite-element methods.  
Both of these approaches 
deal with discrete-time systems that, in the classical limit, generically exhibit chaos.
 
Integrable classical mappings, as opposed to  
mappings that exhibit chaos,  
have been systematically constructed and studied in recent years, see, for example, \cite{Gram:dpe},
\cite{qrt:qrt1},\cite{qrt:qrt2},\cite{Suris:standard},\cite{Ves:IntM},\cite{Ves:WhatIs}.
A specific set of examples are the so-called 
mappings of KdV type \cite{Ca:complete}, \cite {PaNi:IntMap}, which were classically constructed 
from periodic initial 
value problems on the lattice KdV \cite{NiQuCa:direct} partial difference equation.  They are the
principal model of interest in this paper.
Integrable mappings
are highly nontrivial;  
however, there is a great deal of understanding of them, both  
on the level of their  
solvability as well as on their classification.  
 
Historically the quantization of integrable models formed many of 
the paradigms of quantum theory.  Furthermore, the inherent discreteness of quantum theory points 
to a specific r\^ole that discrete,  
and hence integrable discrete, systems may play in the further development of the  
theory.
Quantum integrability on the discrete space-time lattice has been considered in a number of papers
(mostly from the  
perspective of $R$-matrix theory), for example, 
\cite{Ca:IntQM}, \cite{fadvol:aqoimidst}, \cite{NiCa:Quantof}, \cite{NiCaPa:IntQuant},
\cite{Resh:ISWDT}, \cite{vol:quant}.
Here the discrete aspect is not seen as some kind of approximation but, rather, it is  
postulated as the  
underlying structure of space-time from the very start.  
In the present paper, adopting the same point of view, we will be mainly  
concerned with the approach  
initiated in 
\cite{Ca:IntQM}, \cite{NiCa:Quantof}, \cite{NiCaPa:IntQuant}, where a 
non-ultralocal Yang-Baxter ($R$-matrix) structure appropriate for obtaining an 
``integrable quantization'' of the mappings of KdV type was given
(in the continuous-time setting such a non-ultralocal Yang-Baxter structure
had been previously used in connection with the quantum Toda theory \cite{BaBo:QTT}).
 
Quantizing a  
discrete-time system 
is essentially different to the conventional quantization procedure of Hamiltonian systems.  In the  
continuous-time setting it 
is the Hamiltonian (or commuting family of Hamiltonians) that stand central to the theory, the 
spectrum and  
eigenfunctions of which  
are the main objects to be computed. The equations of motion in terms of the  
canonical operators  
(in the Heisenberg picture), or the evolution of the states (in the  
Schr\"odinger picture), only play a  
subsidiary r\^ole. In the discrete-time setting it is no longer the  
Hamiltonian(s) that define the model,  
but the equations of motion exclusively. This draws us away from the  
conventional schemes of quantization,  
and leads us to investigate more closely the quantum equations of motion.  
Following the historical  
imperative once again, integrability offers a leading principle to develop our understanding of quantum  
mechanics, this time in the  
discrete regime.

By an  
integrable quantum mapping we mean an automorphism of the  
quantum algebra under consideration, which, furthermore, possesses a 
``sufficient'' set of  
exact commuting invariant functions on the algebra.  
(We will make this definition more precise later.)
This is  
the quantum counterpart of the classical  
integrability of mappings in the sense of Liouville-Arnol'd-Veselov \cite{Ves:IntM}. On the classical level,  
mappings of this type typically involve rational expressions exhibiting  
singularities that imply that the time-evolution  
cannot be globally defined. The characterization of  
integrable maps through their singularity structure  
is a focus point of current investigation, see, for example, \cite{GrRaPa:PP},  
\cite{Sakai:geomPainleve}. 
In the more restrictive context of the mappings  
of KdV type, 
a natural resolution is  
to describe the mapping in terms of variables that live on the  
Riemann surface associated with the underlying spectral curve.  The explicit  
description of the dynamics on the Riemann  
surface was achieved in \cite{Ni:discdub} (see also \cite{EnLeNi:Abel}, \cite{NiEn:Intmaphyp}) 
leading to the so-called \textit{discrete Dubrovin  
equations}.

With this insight, one may take the point of view that the proper quantization procedure for this 
discrete-time system is to write the equations of motion in terms of the quantum analogue of these 
variables.  These variables are Sklyanin's variables (which are also known as quantum separated 
variables).  Following the correspondence with the classical case, the equations of motion in terms of 
the Sklyanin variables are called \emph{quantum discrete Dubrovin equations}.
(It should be noted that the Sklyanin variables are taking an increasingly primary r\^ole in the field of integrable systems.  They 
have played a fundamental part in various recent publications, with motivations different to that of this 
work, such as \cite{Ba:Inverse}, \cite{BaTa:Sov}, \cite{Smir:DualBax}.)

The outline of the paper is as follows.
The non-ultralocal Yang-Baxter structure of \cite{NiCaPa:IntQuant} is recapitulated in section \ref{NULYBS}, in such a way
as to bring the features required for the derivation of the quantum
discrete Dubrovin equations to the fore.
More specific information pertaining
to the mappings of KdV type is given in section \ref{qmkdv}.
In section \ref{SV} the Sklyanin algebra is set up, this paves the
way for the derivation of the quantum discrete Dubrovin equations in section \ref{qdde}.
(Equations (\ref{qddmatrixskly}) and (\ref{xupsym}) give a temporal, that is discrete-time,
evolution and, hence, are called \emph{temporal quantum discrete Dubrovin equations.}
Equations (\ref{SQDD}) and (\ref{xspaceupsym}) give a spatial evolution and, hence, are
called \emph{spatial quantum discrete Dubrovin equations}.)
The reconstruction of the original dynamical variables (of the mappings of KdV type)
in terms of the Sklyanin variables is also addressed in section \ref{qdde}.
The well-defined evolution arising from the quantum discrete Dubrovin equations and the
reconstruction is illustrated in the one and two degrees of freedom situations in section \ref{exs}.
We remain very formal and algebraic throughout this paper, principally concentrating on the derivation 
of the quantum discrete Dubrovin equations.

	\section{Non-ultralocal Yang-Baxter structure} \label{NULYBS}

The non-ultralocal Yang-Baxter structure 
for the class of discrete-time systems to which the mappings of KdV type belong
was given in \cite{NiCaPa:IntQuant}.  The convention
of that paper (the standard convention) will be employed in this section, 
\ref{invproof}, and \ref{qdetappan}.  This convention
includes that the subscripts $1, 2, \ldots$ denote factors in a matricial tensor product, and the
same subscripts distinguish the associated spectral parameters.
Care must be taken not to confuse these subscripts with those that correspond to the grading of the 
monodromy matrix or the subscripts which identify different dynamical
variables.  These are, however, all perfectly clear within their context.

The  only nontrivial commutation relations between 
the operators $L_n(\lambda )$ are
those on 
the same and nearest-neighbour sites, namely as follows

\numparts 
\bea
R_{12}^+\,L_{n,1} \,L_{n,2} &=& 
L_{n,2} \,L_{n,1}\,R_{12}^-\  ,  \label{eq:RLL}  \\
L_{n+1,1}  S^+_{12}\,L_{n,2} &=& 
L_{n,2} \,L_{n+1,1}\   ,  \label{eq:LSL}  \\
L_{n,1}  \, L_{m,2} &=& L_{m,2}  \, L_{n,1} \qquad \mid n-m \mid 
\geq 2\    , \label{eq:2.1c}
\eea\endnumparts 
where $L_n(\lambda )$ is the $L$ operator at the nth site and $L_{n,j}$
denotes $L_n(\lambda )$ acting nontrivially only on the jth factor of the tensor product,

	\begin{displaymath}
	L_{n,j} := {\bf 1}\otimes{\bf 1}\otimes \ldots \otimes \underbrace{L_n(\lambda_j )}_{\textrm{jth place}}
	\otimes \ldots \otimes{\bf 1}.
	\end{displaymath}
The operators $R^{\pm}_{j k} := R^{\pm}_{j k}(\lambda_j, \lambda_k)$ act nontrivially only on
the jth and kth factors of the tensor product.
As was shown in \cite{NiCaPa:IntQuant}, equations (\ref{eq:RLL}) to (\ref{eq:2.1c}) lead to

\numparts
    \bea
R^{+}_{12}\,T^{\pm}_{n,1}\,T^{\pm}_{n,2} &=& 
T^{\pm}_{n,2}\,T^{\pm}_{n,1}\,R^{-}_{12}\  ,  \label{eq:RTT} \\
T^{+}_{n,1}\,S^{+}_{12}\,T^{-}_{n,2} &=& 
T^{-}_{n,2}\,S^{-}_{12}\,T^{+}_{n,1}\  , \label{eq:TST} 
	\eea\endnumparts
where $S^+_{12}=S^-_{21}$,
\be
T^{+}_n(\lambda )\  \ :=  \ \stackrel{\longleftarrow}{\prod_{j=n+1}^P}\ 
L_j(\lambda )    \qquad  T^{-}_n(\lambda )\  \ :=  \ \stackrel{\longleftarrow}{\prod_{j=1}^n}\ 
L_j(\lambda )\ ,	\label{Tplusminus}
\ee
($P \in \mathbb{N}$ denotes the ``spatial'' periodicity of the model, $1 \le n \le P-1$)
and also
lead to
\be
R_{12}^{+}\,T_1 \,S^{+}_{12}\, T_2 \ =\ T_2 \, S^{-}_{12}\,
T_1\, R_{12}^{-}\    , \label{RTST}
\ee
where $T_1$ denotes the monodromy matrix, $T(\lambda)$, acting
nontrivially only on the first factor of the tensor product; the monodromy matrix, 

\be
T(\lambda )\  :=  \ \stackrel{\longleftarrow}{\prod_{n=1}^P}\ 
L_n(\lambda )\  . \label{T}
\ee

\noindent
(The convention for the ordered product in (\ref{Tplusminus}) and (\ref{T})
is such that the matrices $L_n$ are ordered from right to left with increasing label $n$.)  
In the $P=2$ case equation (\ref{eq:TST}) replaces (\ref{eq:LSL}).
The compatibility relations of equations (\ref{eq:RLL}) to (\ref{eq:2.1c}) lead 
to the following consistency conditions on $R^{\pm}$ and $S^{\pm}$: 
 \numparts   
\bea
R^{\pm}_{12}\,R^{\pm}_{13}\,R^{\pm}_{23} &=& 
R^{\pm}_{23}\,R^{\pm}_{13}\,R^{\pm}_{12}\  ,  \label{eq:RRR} \\
R^{\pm}_{23}\,S^{\pm}_{12}\,S^{\pm}_{13} &=& 
S^{\pm}_{13}\,S^{\pm}_{12}\,R^{\pm}_{23}\  . \label{eq:RSS} 
\eea\endnumparts
Equation (\ref{eq:RRR}) is the quantum Yang-Baxter equation 
for $R^{\pm}$, which is coupled with  
$S^{\pm}$ by equation (\ref{eq:RSS}).
It is also assumed that $S^-_{12}$ and $S^+_{12}$ are invertible.
In order to establish that the structure given by the above commutation relations 
allows for suitable commutation relations for the monodromy 
matrix we need to impose in addition to (\ref{eq:RRR}) and (\ref{eq:RSS}) that
\be
R_{12}^{\pm}\,S^{\pm}_{12}\ =\ S^{\mp}_{12}\,R_{12}^{\mp}\    . \label{RSSR}
\ee

Integrable mappings follow from a discrete-time Zakharov-Shabat
system of the form
	\begin{equation}
	\widetilde{L}_n(\lambda) M_n(\lambda) = M_{n+1}(\lambda) L_n(\lambda),	\label{ZS}
	\end{equation}
where the tilde, $\widetilde{ \,} $ , denotes a time update and $M_n(\lambda)$ is the discrete-time evolution
operator at the site n ($\widetilde{M}_n(\lambda)$ would denote the
discrete-time evolution operator at the site n at the next time level).
The M (or temporal) part of the extended Yang-Baxter structure, as given in \cite{Ca:IntQM}, \cite{NiCa:Quantof},
and \cite{NiCaPa:IntQuant}, allows one to derive the invariants of the discrete-time evolution.  
It also
allows one to show that the Yang-Baxter relation (\ref{RTST}) is preserved throughout the
(discrete) time evolution.  The only extra
relations from the M part of the extended Yang-Baxter structure required for these proofs are

\be
R^+_{12}\,M_{n,1} M_{n,2} = M_{n,2} M_{n,1} R^-_{12} \  \label{eq:RMM}
\ee
and 

\be
T_1 \, M_{1,1}^{-1} \,S^{+}_{12}\, M_{1,2} \ =\ M_{1,2} \, S^{-}_{12}\,
T_1\, M_{1,1}^{-1}\  .   \label{TMSM}
\ee

\noindent
The proofs are given in \ref{invproof}.

In the quantum discrete-time setting the commuting family of invariants (i.e., the invariants of the discrete-time evolution)
are given by expanding
\be  
\tau( \lambda )\ =\ \tr\left( K(\lambda )T(\lambda )\right)\       \label{trinv}
\ee
in powers of the spectral parameter, $\lambda$.  The proof follows by taking the trace 
over both spaces of the tensor product
of $P_{12}K_1K_2$ multiplying
equation (\ref{TMSM}) from the left 
(where $P_{12}$ is the permutation operator).  
(The details can be found in \ref{invproof}.)
The result is that

\be \label{eq:K} 
K_2\,=\,\tr_1\left\{ P_{12} \,^{t_1\!}\left[ ( \,^{t_1\!}S^+_{12})^{-1}
\right] \right\} \    , 
\ee 
where the left superscript $^{t_1\!}$ denotes the matrix transpose in the first factor of the
matricial tensor product.  Observe that the commutation relation for the $L_n$ operators,
equation (\ref{eq:RLL}), is of the same form as that for the $M_n$ operators,
equation (\ref{eq:RMM}), and it follows immediately from equation (\ref{eq:TST}) that

	\begin{equation}
		 T_1 \, L_{1,1}^{-1} \,S^{+}_{12}\, L_{1,2} \ =\ L_{1,2} \, S^{-}_{12}\,
		T_1\, L_{1,1}^{-1}\  ,   \label{TLinv}
	\end{equation}
which is of the same form as equation (\ref{TMSM}).
Hence it follows, in an exactly analogous fashion to the temporal evolution,
that there is a ``spatial'' evolution which preserves the
Yang-Baxter relation (\ref{RTST}) and the family of invariants (\ref{trinv}).
This is proven (in both the spatial and temporal case) in
\ref{invproof}. 
The spatial evolution is denoted by the hat, $\widehat{ \,} $ , 
hence

\be
	\widehat{T}(\lambda) = L_1(\lambda) T(\lambda) L_1(\lambda)^{-1}. \label{Spatup}
\ee

\noindent

For later purposes it is now assumed that $R^-_{12}$ is proportional to a rank-one projector
for a particular relative value of the spectral parameters $\lambda_1$ and $\lambda_2$.
This occurs for a number of quantum models \cite{korepin:qismacf}.  From equations (\ref{RTST})
and (\ref{RSSR}),

\be
R_{12}^{-}\, S_{12}^{-}\, T_1 \,S^{+}_{12}\, T_2 \ =\ S^{+}_{12}\, T_2 \, S^{-}_{12}\,
T_1\, R_{12}^{-}\    . \label{SRTST}
\ee
Assuming the particular relative value of $\lambda_1$ and $\lambda_2$ such that 
$R^-_{12}$ is proportional to a rank-one projector, the quantum determinant \cite{ReKuSk:81}, \cite{KuSk:82},
is denoted by $\Delta$, where

\be
R^-_{12} \, \Delta \ =\ R_{12}^{-}\, S_{12}^{-}\, T_1 \,S^{+}_{12}\, T_2    . \label{RTSTdet}
\ee
{}From equations (\ref{eq:RLL}) and (\ref{RSSR})

\be
R_{12}^- S^{-}_{12}\,L_{n,1} \,L_{n,2} = 
S^{+}_{12} L_{n,2} \,L_{n,1}\,R_{12}^-\ . \label{RSLL}
\ee
Maintaining the same particular relative value of $\lambda_1$ and $\lambda_2$,
the local quantum determinant is denoted by $\textrm{Qet}(L_n)$, where

\be
R_{12}^- \textrm{Qet}(L_n) = 
R_{12}^- S^{-}_{12}\,L_{n,1} \,L_{n,2} . \label{RLLdet}
\ee
In \ref{qdetappan} it is shown that the quantum determinant factorizes in terms
of the local quantum determinants as

\be
\Delta = \stackrel{\longleftarrow}{\prod_{n=1}^{P}}\ 
\textrm{Qet}(L_n).  \label{deltaproductfull}
\ee
In section \ref{qmkdv} the quantum mappings of KdV type are considered.
The quantum determinant and the local quantum determinants
are central elements of the algebra for this model.
Indeed, the quantum determinant will play a central r\^ole,
in both the mathematical and conventional English sense, throughout the rest of this paper.

	\section{Quantum mappings of KdV type}\label{qmkdv}

In operator form the quantum mappings of KdV type, as introduced in \cite{NiCa:Quantof} and \cite{NiCaPa:IntQuant}, are

\be 
\widetilde{v}_{2j-1}\ =\ v_{2j}   \qquad  \widetilde{v}_{2j}\ =\ v_{2j+1}\ -\ a {v_{2j}}^{-1}\ 
+\ a{v_{2j+2}}^{-1}  \qquad (j=1,\cdots ,
P),  \label{eq:xy}
\ee

\noindent
with imposed periodicity condition $v_{i+2P}=v_i$, $P \in \mathbb{N}$.
The dynamical variables, $v_n$, are Hermitian operators, $a$ is a real number parameter.  In the
notation of \cite{NiCa:Quantof} the commutation relations of the
dynamical variables read

\be
\left[v_j\,,\,v_{j'} \right]\ =\ h\,\left(\dd_{j,j'+1}\,-
\,\dd_{j+1,j'}\right)\    ,  \label{eq:xycr}
\ee
($h = - i \hbar$, where $\hbar$ is Planck's constant divided by $2 \pi$).
The periodic
initial value problem, from which the mapping arose, imposes on the mapping 
(\ref{eq:xy}) the Casimirs 
\be
\sum_{j=1}^{P} v_{2j}\,=\,\sum_{j=1}^{P} v_{2j-1}\,=:\,\nu \  , \label{eq:4.2}
\ee
in such a way as to leave the value of these Casimir operators as a free parameter
(it can easily be seen from the commutation relation, equation (\ref{eq:xycr}), that this is a Casimir) hence, in the classical limit,
we obtain what could be called a $(P - 1)$-dimensional 
configuration space generalization of the McMillan 
mapping \cite{McMillan:map}. We assume that $\nu \neq 0$. 

We need to point out that, at this stage, we are only concerned with the
algebraic structures behind the integrability of the quantum discrete-time
systems. Hence, as far as this paper is concerned, we will deal with operators,
such as the $\{ v_{k} \} := \{ v_{k} \}_{k=1\ldots 2P}$, on a strictly formal level (in the spirit of related
work \cite{fadvol:aqoimidst}).  This involves, for instance, assumptions on the invertibility
of the operators, disregarding, for the time being, questions concerning the
domains of the Hilbert spaces on which they act (we aim to return to this
latter issue in subsequent publications).

The Lax description of the mappings of KdV type is as follows:	
\be
L_j\ =\ V_{2j} \, V_{2j-1} \qquad
 V_i\ =\ \left( \begin{array}{cc} v_i&1\\ 
\lambda_i&0 \end{array} \right) \label{L}
\ee
and $\lambda_{2j}=\lambda$, $\lambda_{2j+1} = \lambda + a$. 
The associated monodromy matrix, $T(\lambda )$, is
obtained by gluing the elementary translation matrices $L_j$ along a line 
connecting the sites 1 and $P+1$ over one period $P$, namely
\be
T(\lambda )\ =\ \left( \begin{array}{cc} A(\lambda)&B(\lambda)\\ 
C(\lambda)&D(\lambda) \end{array} \right) \ :=  \ \stackrel{\longleftarrow}{\prod_{n=1}^P}\ 
L_n(\lambda )\  . \label{Tmatrix}
\ee

\noindent
The monodromy matrix has a natural grading in terms of the spectral parameter, $\lambda$,
\be
\fl T(\lambda )\ =\ \left( \begin{array}{cc} \lambda^{P} + \lambda^{P-1} A_{P-1} + \ldots + A_0 
		& \lambda^{P-1} B_{P-1} + \lambda^{P-2} B_{P-2} + \ldots + B_0 \\ 
\lambda^{P} C_{P} + \lambda^{P-1} C_{P-1} + \ldots + \lambda C_1   
		& \lambda^{P} + \lambda^{P-1} D_{P-1} + \ldots + \lambda D_1 \end{array} \right) \ . \label{Tgraded}
\ee
Observe that $A(\lambda)$ and $D(\lambda)$ are both monic polynomials in $\lambda$.
The time evolution is given by
\be
\widetilde{T}(\lambda) = M(\lambda) T(\lambda) M(\lambda)^{-1}   \qquad
M_n\ =\ \left( \begin{array}{cc} w_n&1\\ 
\lambda&0 \end{array} \right), \label{evolve}
\ee
where $M$ is $M_1$, the discrete-time evolution operator at lattice site 1.
More explicitly, bearing in mind that we are dealing with noncommuting operators, this gives us,
\be
\widetilde{T}(\lambda )\ =\ \left( \begin{array}{cc} w B + D&\frac{1}{\lambda}\left(w A + C - w B w - D w   \right)\\ 
\lambda B&A - B w \end{array} \right) , \label{UPDATE}
\ee
where $w := w_1$.  As well as the mapping (\ref{eq:xy}), the Zakharov-Shabat condition (\ref{ZS}) reveals that

	\begin{equation}
	w_n = v_{2n-1} + \frac{a}{v_{2n}}. \label{wasvees}
	\end{equation}

For the mappings of KdV type, the realization of the $R$ and $S$ matrices, which are solutions of 
the compatibility relations (\ref{eq:RRR}) and (\ref{eq:RSS})
under the condition (\ref{RSSR}), is as follows:
\bea
R^+_{12} & = & R^-_{12} - S^+_{12} + S^-_{12} \    \nn \\
R^-_{12} & = & {\bf 1}\otimes {\bf 1}\ +\ 
h\,\frac{P_{12}}{\lambda_1\,-\,\lambda_2}\  \label{eq:Rsol} \\
S^+_{12} & = & {\bf 1}\otimes {\bf 1}\ - \frac{h}{\lambda_2} F \otimes E \qquad  S^-_{12} = S^+_{21} \  , \nn
\eea
where the permutation 
operator $P_{12}$ and the matrices $E$ and $F$ are given by
\be 
P_{12} \ =\ \left( \begin{array}{cccc} 
1&0&0&0\\0&0&1&0\\0&1&0&0\\0&0&0&1 \end{array} \right) \qquad
E\ =\ \left( \begin{array}{cc} 
0&1\\0&0 \end{array} \right) \qquad F\ =\ 
\left( \begin{array}{cc} 
0&0\\1&0 \end{array} \right) \  .     \label{eq:4.14}
\ee
The realization (\ref{eq:Rsol}) is assumed throughout the rest of the paper.

Classically equation (\ref{evolve}) gives us that the trace of the monodromy matrix is invariant
under the discrete-time evolution.  This argument no longer holds in the quantum case, as some of the matrix entries consist of
noncommuting operators.
As stated in the previous section, within the quantum case the invariants 
are given by expanding
equation (\ref{trinv}) 
in powers of the spectral parameter, $\lambda$.  For mappings of KdV type
equation (\ref{eq:K}) gives
\be
K(\lambda )\ =\ \left( \begin{array}{cc} 1&0\\ 
0&1 + \frac{h}{\lambda} \end{array} \right) \  
\ee
thus
\be  
\tau( \lambda )\ =\ A(\lambda) + \left( 1 + \frac{h}{\lambda}\right) D(\lambda)  .     \label{ADinv}
\ee

Equation (\ref{eq:Rsol}) shows that $R_{12}^-$ is 
the fully antisymmetric
projector, ${\bf1} - P_{12}$, when $\lambda_2 = \lambda_1 + h$.
Therefore, with $\lambda_1 = \lambda - \frac{h}{2} =: \lambda_{-}$ 
and $\lambda_2 = \lambda + \frac{h}{2} =: \lambda_{+}$, equation (\ref{RTSTdet})
gives the quantum determinant for this model.  It may be expressed as

\be
\Delta(\lambda) \ =\ \frac{\lambda_{+}}{\lambda_{-}} D(\lambda_{-})
A(\lambda_{+}) - B(\lambda_{-}) C(\lambda_{+}). 
\label{Delta1}
\ee

\noindent
There are, of course, other equivalent expressions which can be obtained using 
the algebra (\ref{RTST}), for instance
\be
\Delta(\lambda) \ =\ \frac{\lambda_{+}}{\lambda_{-}} A(\lambda_{+})
D(\lambda_{-}) 
- \frac{\lambda_{+}}{\lambda_{-}}
B(\lambda_{+}) C(\lambda_{-}).    \label{Delta2}
\ee
Similarly, if we write
\be
L_n(\lambda )\ =\ \left( \begin{array}{cc} a_n(\lambda)&b_n(\lambda)\\ 
c_n(\lambda)&d_n(\lambda) \end{array} \right),
\ee
then it can easily be shown using (\ref{RLLdet}) that the quantum determinant of the algebra (\ref{eq:RLL})
can be written as,

	\be
\textrm{Qet}(L_n(\lambda)) \ =\ \frac{\lambda_{+}}{\lambda_{-}} d_n(\lambda_{-})
a_n(\lambda_{+}) - b_n(\lambda_{-}) c_n(\lambda_{+}). \label{smalldet}
	\ee
Hence, from (\ref{L}), $\textrm{Qet}(L_n(\lambda)) = \lambda_+(\lambda_+ + a)$ and, therefore,
from (\ref{deltaproductfull}),

	\begin{equation}
	\Delta(\lambda) = \lambda_+^P(\lambda_+ + a)^P, \label{kdvdet}
	\end{equation}
which manifestly belongs to the centre of the algebra.

	\section{Sklyanin variables}\label{SV}

Following Sklyanin \cite{Sklyanin:qism} the operator zeros of $B(\lambda)$, $x_n$, 
provide the separated canonical variables.  
By ``operator zeros'' it is meant that
\be
B(\lambda )\ =\  B_{P-1}\prod_{n=1}^{P-1}\ 
(\lambda - x_n)\  , \label{opzero}
\ee
where the $\{ x_i \} := \{ x_i \}_{i=1 \ldots P-1}$ mutually commute.
(In \ref{BCconstant} it is shown that $B_{P-1}$ is equal to the Casimir (\ref{eq:4.2}).
We assume the mutual commutativity of the $\{ x_i \}$, this is consistent with equation (\ref{RTST}).)
Conjugate variables to the $x_n$ are 
introduced by making the definitions
\be
X_n^{-} = A(\lambda) \bigg |_{\lambda = x_n}  \qquad
X_n^{+} = \left( 1 + \frac{h}{\lambda} \right) D(\lambda) \bigg |_{\lambda = x_n} \label{XX}
\ee

\noindent
where the operator ordering prescription throughout this paper is that $x_n$ is substituted 
for the spectral parameter, $\lambda$, from the left, thus,
\numparts
\be 
	X^-_n = A_0 + x_n A_1 + \ldots + x_n^P,\label{minusgrade}
\ee
\be
	X^+_n = h D_1 + x_n (D_1 + h D_2) + \ldots + x_n^P.\label{plusgrade}
\ee
\endnumparts

\noindent
As in \cite{Sklyanin:qism} the full set of commutation relations between these operators
follows from the Yang-Baxter structure, equation (\ref{RTST}), and reads,

\numparts
\be
[  x_m \,, \, x_n ] = 0 \qquad \forall \, m , n  \label{com1}
\ee
\be
X^{\pm}_m x_n = (x_n \pm h \, \delta_{mn}) X^{\pm}_m \qquad \forall \, m , n
\label{com2}
\ee
\be
[  X^{\pm}_m \,,\,  X^{\pm}_n ] = 0 \qquad \forall \, m , n
\label{com3}
\ee
\be
[  X^{-}_m \,,\,  X^{+}_n ] = 0 \qquad \forall \, m \neq n
\label{com4}
\ee
\be
X^{\pm}_n X^{\mp}_n = \Delta\left(x_n \pm \frac{h}{2} \right) \qquad \forall \,
n \label{comlast}
\ee
where $\Delta(\lambda)$ is the quantum determinant of the model,
given explicitly in equation (\ref{kdvdet}).
We also have
\be
\tau(x_n) = X_n^+ + X_n^-, \label{Bax1}
\ee 
\endnumparts
which leads to the linear finite-difference spectral problem known as Baxter's equation (see, for example, 
\cite{Sklyanin:qism} and \cite{Sklyanin:sov}).

The derivation of the Sklyanin algebra relations
(equations (\ref{com1}) to (\ref{Bax1})) proceeds in the same way as in
\cite{Sklyanin:qism}, but is slightly more involved as the initial equations
from the Yang-Baxter equation are more complicated (there are extra terms due, essentially, to the
non-ultralocal nature of this algebra).  Remarkably, as the derivation
is carried out, the extra terms vanish, leaving the Sklyanin algebra relations.

The proof of the preservation of equation (\ref{RTST}),
for both the temporal and spatial evolutions, is given in \ref{invproof}.
As a consequence we have the following result for the Sklyanin algebra
under these evolutions.

\vspace{5mm}

\noindent
\textbf{Proposition.} \emph{The Sklyanin algebra relations are preserved
under both the temporal and spatial discrete evolutions.}

\vspace{5mm}

The (extended) Yang-Baxter structure of section \ref{NULYBS} leads efficiently
to the preservation of the Sklyanin algebra relations, as is expressed in the proposition.
The Sklyanin algebra relations are the real starting point of this work.  We now
turn to the equations of motion in terms of the Sklyanin algebra variables.

	\section{Quantum discrete Dubrovin equations}\label{qdde}

The equations of motion, for the temporal and the spatial evolutions,
are derived in this section.  The aim is to establish these discrete evolutions
in terms of the Sklyanin algebra variables; this requires the reconstruction of $w$ and one of the
original dynamical variables, $1/v_2$, in terms of the Sklyanin algebra variables.
Hence the issue of the reconstruction of the original dynamical variables in terms of the Sklyanin algebra
variables is also, necessarily, addressed.

In section \ref{qddeinv} the invariants of both the temporal and spatial evolutions are expressed
in terms of the Sklyanin algebra variables.

In section \ref{qddte} the reconstruction of $w$ in terms of the Sklyanin algebra variables is given.
This paves the way for the derivation of equations (\ref{qddmatrixskly}) and (\ref{xupsym}).
As is illustrated for the $P=2$ and $P=3$ cases in section \ref{exs}, these two equations, along with the
preservation of the invariants, give a well-defined temporal evolution.
Hence, equations (\ref{qddmatrixskly}) and (\ref{xupsym}) of section \ref{qddte} are what we mean by
the \emph{temporal quantum discrete Dubrovin equations}.

In section \ref{spatialsect} the reconstruction of $1/v_2$ in terms of
the Sklyanin algebra variables is given.  Along with the reconstruction of $w$
this paves the way for the derivation of equations (\ref{SQDD}) and (\ref{xspaceupsym}).
In an exactly analogous fashion to (\ref{qddmatrixskly}) and (\ref{xupsym}) in the temporal case,
equations (\ref{SQDD}) and (\ref{xspaceupsym}) give a spatial evolution. 
Hence, equations (\ref{SQDD}) and (\ref{xspaceupsym}) of section \ref{spatialsect} are what we mean by the
\emph{spatial quantum discrete Dubrovin equations}. 
We conjecture that the spatial
evolution allows for a reconstruction of all of the original dynamical variables in terms of the unshifted Sklyanin
variables.
This is illustrated in the $P=3$ case in section \ref{exs}, and it is not technically difficult
to confirm this for the next few larger-period cases.  However, the calculations quickly become
very cumbersome as the period increases.

	\subsection{Invariants}\label{qddeinv}

In this section expressions are given for the invariants, which are the coefficients of the
various powers of $\lambda$ in equation (\ref{ADinv}), in terms of the Sklyanin algebra variables.
{}From the form of the invariant in terms of entries of the monodromy matrix, (\ref{ADinv}), and 
their gradation (\ref{Tgraded}),

	\be
	\tau(x_n) = 2x_n^P + x_n^{P-1}I_{P-1} + x_n^{P-2}I_{P-2} +
	\ldots + x_nI_1 + I_0.\label{tauofx}
	\ee
It is easily shown that $I_{P-1}$ is a Casimir operator.  Taking into account the gradation given in equation (\ref{Tgraded}),
equation (\ref{Delta1}) gives us that the quantum determinant

	\begin{displaymath}
	\Delta(\lambda + \frac{h}{2}) = \lambda^{2P} + \lambda^{2P-1} ( Ph + A_{P-1} + D_{P-1} + h - B_{P-1}C_P ) + \ldots.
	\end{displaymath}
However, for the
mappings of KdV type we have (from equation (\ref{kdvdet})),

	\begin{displaymath}
	\Delta(\lambda + \frac{h}{2}) = \lambda^{2P} + \lambda^{2P-1} ( 2Ph + Pa ) + \ldots.
	\end{displaymath}
Therefore, using that the central elements $B_{P-1} = C_P = \nu$ (as derived in \ref{BCconstant}),
\be
I_{P-1} \ :=\ A_{P-1} + D_{P-1} + h \ =\ \nu^2 + P (a + h) .\label{IPminus1}
\ee

\noindent	
Now observe that,

\be
	\left( \begin{array}{c} 
	I_0\\ 
	I_1\\ 
	\vdots \\
	I_{P-2} \end{array} \right) \,
	= \,  
	\left( \begin{array}{cccc} 
	1&x_1&\ldots&x_1^{P-2}\\ 
	1&x_2&\ldots&x_2^{P-2}\\ 
	\vdots& & &\vdots \\
	1&x_{P-1}&\ldots&x_{P-1}^{P-2} \end{array} \right)^{-1}
	\left( \begin{array}{c} 
	\tau(x_1) - 2x^P_1 - x_1^{P-1}I_{P-1}\\ 
	\tau(x_2) - 2x^P_2 - x_2^{P-1}I_{P-1}  \\ 
	\vdots \\
	\tau(x_{P-1}) - 2x^P_{P-1} - x_{P-1}^{P-1}I_{P-1}   \end{array} \right) .\label{taunoughtfmla}
\ee
All terms on the right hand side are known 
(remember that $\tau(x_n) = X_n^+ + X_n^-$).
Equation (\ref{tauofx}) may also be rewritten as

\be
	\left( \begin{array}{c}  
	I_1\\ 
	I_2\\ 
	\vdots \\
	I_{P-1} \end{array} \right) \,
	= \,  
	\left( \begin{array}{cccc} 
	x_1&x_1^2&\ldots&x_1^{P-1}\\ 
	x_2&x_2^2&\ldots&x_2^{P-1}\\ 
	\vdots& & &\vdots \\
	x_{P-1}&x_{P-1}^2&\ldots&x_{P-1}^{P-1} \end{array} \right)^{-1}
	\left( \begin{array}{c} 
	\tau(x_1) - 2x^P_1 - I_0\\ 
	\tau(x_2) - 2x^P_2 - I_0\\ 
	\vdots \\
	\tau(x_{P-1}) - 2x^P_{P-1} - I_0   \end{array} \right) .\label{invcloumn}
\ee
Note that within the VanderMonde matrices all of the entries commute.

As the quantum discrete Dubrovin equations, which are derived in the following sections,
are also matricial equations, it is expedient to introduce some specialized notation.
The symbol $\mathcal{M}$ is introduced to denote the VanderMonde matrix,
the symbol $\mathcal{D}$ 
to denote the diagonal matrix,
\begin{displaymath}
	\mathcal{M}
	= \left( \begin{array}{cccc} 
	x_1&x_1^2&\ldots&x_1^{P-1}\\ 
	x_2&x_2^2&\ldots&x_2^{P-1}\\ 
	\vdots& & &\vdots \\
	x_{P-1}&x_{P-1}^2&\ldots&x_{P-1}^{P-1} \end{array} \right) \qquad
	\mathcal{D}
	= \left( \begin{array}{cccc} 
	x_1&0&\ldots&0\\ 
	0&x_2&\ldots&0\\ 
	\vdots& & &\vdots \\
	0&0&\ldots&x_{P-1} \end{array} \right).
\end{displaymath}
Vectors are indicated by a bold typeface, and all are $P-1$ dimensional.  
The vector $\mathbf{e} = (1,1, \ldots,1)^t$.
The vector consisting of ordered entries, labelled $n$ to $P - 2 + n$, for any integer $n$, is
denoted by a bold typeface with a subscript $n$, for instance
$\mathbf{y}_0 = (y_0,y_1, \ldots,y_{P-2})^t$, $\mathbf{y}_1 = (y_1,y_2, \ldots,y_{P-1})^t$.
In this notation equations (\ref{taunoughtfmla}) and (\ref{invcloumn}) are 
rewritten as

	\begin{displaymath}
	\mathbf{I}_0 = \mathcal{M}^{-1} \mathcal{D} \left(\mathbf{X}^+_1 + \mathbf{X}^-_1 - 
	\mathcal{D}^P 2 \mathbf{e} - \mathcal{D}^{P-1} I_{P-1} \mathbf{e}   \right)
	\end{displaymath}
and

	\begin{displaymath}
	\mathbf{I}_1 = \mathcal{M}^{-1} \left(\mathbf{X}^+_1 + \mathbf{X}^-_1 - 
	\mathcal{D}^P 2 \mathbf{e} - I_{0} \mathbf{e}   \right),
	\end{displaymath}
respectively, where $\mathbf{I}_0$ and $\mathbf{I}_1$ denote the left-hand sides
of (\ref{taunoughtfmla}) and (\ref{invcloumn}) respectively.

	\subsection{Temporal equations}\label{qddte}

In section \ref{Rectime} we give the reconstruction of $w$ in terms of the Sklyanin algebra
variables.  For heuristic reasons the matricial equations are written out in full in this section.
In section \ref{tqdde} the temporal quantum discrete Dubrovin equations are derived.

		\subsubsection{Reconstruction of $w$.}\label{Rectime}

Observe that, from the definitions (\ref{XX})
(and remembering that $x_n$ are substituted for $\lambda$ from the left),

	\begin{displaymath}
	D(\lambda) \bigg |_{\lambda = x_n} := D(x_n) = \frac{x_n}{x_n+h}  X_n^+.
	\end{displaymath}
(We take the casual attitude of writing $1/Y$ for the quantum inverse of the operator $Y$.)
Hence, still working from the definitions,

\be
\left( \begin{array}{c} 
	A_1\\ 
	A_2\\ 
	\vdots \\
	A_{P-1}\end{array} \right) \,
	= \, \left( \begin{array}{cccc} 
	x_1&x_1^2&\ldots&x_1^{P-1}\\ 
	x_2&x_2^2&\ldots&x_2^{P-1}\\ 
	\vdots& & &\vdots \\
	x_{P-1}&x_{P-1}^2&\ldots&x_{P-1}^{P-1} \end{array} \right)^{-1}
	\left( \begin{array}{c} 
	X_1^- - x_1^P -A_0\\ 
	X_2^- - x_2^P -A_0\\ 
	\vdots \\
	X_{P-1}^- - x_{P-1}^P -A_0  \end{array} \right), \label{A}
\ee
\be
\left( \begin{array}{c} 
	D_1\\ 
	D_2\\ 
	\vdots \\
	D_{P-1}\end{array} \right) \,
	= \, \left( \begin{array}{cccc} 
	x_1&x_1^2&\ldots&x_1^{P-1}\\ 
	x_2&x_2^2&\ldots&x_2^{P-1}\\ 
	\vdots& & &\vdots \\
	x_{P-1}&x_{P-1}^2&\ldots&x_{P-1}^{P-1} \end{array} \right)^{-1}
	\left( \begin{array}{c} 
	\frac{x_1}{h+x_1} X_1^+ - x_1^P\\ 
	\frac{x_2}{h+x_2} X_2^+ - x_2^P\\ 
	\vdots \\
	\frac{x_{P-1}}{h+x_{P-1}} X_{P-1}^+ - x_{P-1}^P  \end{array} \right). \label{D}
\ee
So, it is seen that,

	\begin{eqnarray}
\fl	\left( \begin{array}{c} 
	A_1 - D_1 + h D_2\\ 
	A_2 - D_2 + h D_3\\ 
	\vdots \\
	A_{P-1} - D_{P-1} + hD_P \end{array} \right) \nonumber\\
\lo	= \,  \left( \begin{array}{cccc} 
	x_1&x_1^2&\ldots&x_1^{P-1}\\ 
	x_2&x_2^2&\ldots&x_2^{P-1}\\ 
	\vdots& & &\vdots \\
	x_{P-1}&x_{P-1}^2&\ldots&x_{P-1}^{P-1} \end{array} \right)^{-1}
	\left( \begin{array}{c} 
	X_1^- +  \frac{h-x_1}{h+x_1}  X_1^+ - I_0\\ 
	X_2^- +  \frac{h-x_2}{h+x_2}  X_2^+ - I_0 \\ 
	\vdots \\
	X_{P-1}^- +  \frac{h-x_{P-1}}{h+x_{P-1}}  X_{P-1}^+ - I_0  \end{array} \right). \label{dub1}
	\end{eqnarray}

\noindent
The time-update
of the monodromy matrix, equation (\ref{UPDATE}), and a commutation relation contained within (\ref{TMSM}) give

	\begin{equation}\label{ADup}
	\widetilde{A} = B \, w - \frac{h}{\lambda}( A - B \, w) + \left(1+\frac{h}{\lambda}\right)D 
	\qquad \widetilde{D} = A - B \, w .
	\end{equation}
Substituting $x_n$ for $\lambda$ from the left, and again using the definitions (\ref{opzero}) and (\ref{XX}), gives,

\numparts
	\begin{equation}
		\widetilde{A}_0 + x_n\widetilde{A}_1 + \ldots + x_n^P = - \frac{h}{x_n}X_n^- + X_n^+,
	\end{equation}

	\begin{equation}
	x_n\widetilde{D}_1 + x_n^2\widetilde{D}_2 + \ldots + x_n^P = X_n^-.
	\end{equation}
\endnumparts

\noindent
{}From here we easily obtain (using $\widetilde{A}_0 + h\widetilde{D}_1 = \widetilde{I}_0 = I_0$) that

\be
\fl	\left( \begin{array}{c} 
	\widetilde{A}_1 - \widetilde{D}_1 + h \widetilde{D}_2\\ 
	\widetilde{A}_2 - \widetilde{D}_2 + h \widetilde{D}_3\\ 
	\vdots \\
	\widetilde{A}_{P-1} - \widetilde{D}_{P-1} + h\widetilde{D}_P \end{array} \right) \,
	= \, \left( \begin{array}{cccc} 
	x_1&x_1^2&\ldots&x_1^{P-1}\\ 
	x_2&x_2^2&\ldots&x_2^{P-1}\\ 
	\vdots& & &\vdots \\
	x_{P-1}&x_{P-1}^2&\ldots&x_{P-1}^{P-1} \end{array} \right)^{-1}
	\left( \begin{array}{c} 
	X_1^+ -    X_1^- - I_0\\ 
	X_2^+ -    X_2^- - I_0 \\ 
	\vdots \\
	X_{P-1}^+ -   X_{P-1}^- - I_0  \end{array} \right) . \label{dub2}
\ee

\noindent
Equation (\ref{UPDATE}) also
gives

	\begin{displaymath}
	\widetilde{A}(\lambda) -  \widetilde{D}(\lambda) + A(\lambda) - D(\lambda) 
	= w B(\lambda) + B(\lambda) w.
	\end{displaymath}
	
\noindent
A consideration of the leading term as $\lambda \to \infty$ shows that

	\begin{equation}
\widetilde{A}_{P-1} -  \widetilde{D}_{P-1} + A_{P-1} - D_{P-1} = 2 B_{P-1} w,	\label{wAD}
	\end{equation}

\noindent
as $B_{P-1}$ is actually the
Casimir (\ref{eq:4.2}) (this is shown in \ref{BCconstant}).  Therefore

	\begin{displaymath}
w = \frac{1}{2 \nu}(\widetilde{A}_{P-1} -  \widetilde{D}_{P-1} + A_{P-1} - D_{P-1}).
	\end{displaymath}
Expressions for $\widetilde{A}_{P-1} -  \widetilde{D}_{P-1}$ and $A_{P-1} - D_{P-1}$ in terms of the
Sklyanin algebra variables, $\{x_i, X_i^{\pm} \}:=\{x_i , X_i^{\pm} \}_{i=1\ldots P-1}$, follow from equations (\ref{dub1}) and (\ref{dub2}).
To obtain an expression explicitly in terms of $\{x_i, X_i^{\pm} \}$, $I_0$ may be replaced by using (\ref{invcloumn}); then
Cramer's rule, along with (\ref{IPminus1}) for 
the value of the central element $I_{P-1}$, allows one to deduce
that, in terms of the Sklyanin algebra, 

	\begin{equation}
\fl	w = \frac{1}{\nu}\Bigg[ \nu^2 + P a + (P-1)h 
+ (-1)^P\sum_{n=1}^{P-1}
\Bigg( \prod_{{i=1} \atop{i \neq n}}^{P-1} \frac{1}{x_i - x_n}  \Bigg) \Bigg( 2x_n^{P-1}
	-\frac{1}{h + x_n} X_n^+ - \frac{1}{x_n}X_n^-
	  \Bigg)  
     \Bigg] . 	\label{wasxees}
	\end{equation}

		\subsubsection{Temporal quantum discrete Dubrovin equations.}\label{tqdde}

Equations (\ref{dub1}) and (\ref{dub2}) give, in the notation introduced in section \ref{qddeinv},

\begin{equation}
	\widetilde{\mathcal{M}}^{-1}\Big( \mathbf{\widetilde{X}^-}_1 + (h{\bf 1}+\widetilde{\mathcal{D}})^{-1}
	(h{\bf 1}-\widetilde{\mathcal{D}})\mathbf{\widetilde{X}^+}_1
	- I_0 \mathbf{e}  \Big) = 
	\mathcal{M}^{-1}\Big( \mathbf{X}^+_1 - \mathbf{X}^-_1
	- I_0 \mathbf{e}  \Big). \label{qddmatrix}
\end{equation}
This constitutes the (temporal part of the) quantum discrete Dubrovin equations, as given classically in \cite{Ni:discdub}.  
Note that the time-evolved variables
still obey the relations (\ref{com1}) to (\ref{Bax1}) (as stated in the proposition of section \ref{SV}).
Along with the time-update invariance of (\ref{invcloumn}), (\ref{qddmatrix}) leads to

\begin{equation}
	\widetilde{\mathcal{M}}^{-1}\Big( \widetilde{\mathcal{D}}^P\mathbf{e} - 
	(h{\bf 1}+\widetilde{\mathcal{D}})^{-1}\widetilde{\mathcal{D}}\mathbf{\widetilde{X}^+}_1
	\Big) = 
	\mathcal{M}^{-1}\Big( \mathcal{D}^P\mathbf{e} - \mathbf{X}^-_1
	  \Big). \label{qddmatrixskly}
\end{equation}

The equations for the elementary symmetric polynomials in $\{\tilde{x}_i \}$ 
in terms of $\{x_i , X_i^{\pm} \}$ will now be obtained.
A consideration of equation (\ref{opzero}) shows that these follow immediately from the coefficients of different powers of
$\lambda$ in 
$\widetilde{B}(\lambda)$, that is, from $\{\widetilde{B}_0, \ldots, \widetilde{B}_{P-2} \}$.
Equation (\ref{UPDATE}) and a commutation relation contained within (\ref{TMSM}) give

\be
	\widetilde{B}(\lambda) = \frac{1}{\lambda}\left[ A w - B w^2
			+ \left(1 + \frac{h}{\lambda}\right)(C - D w) \right]. \label{Bcomup}
\ee
Hence, to obtain the elementary symmetric polynomials in $\{\tilde{x}_i \}$ in terms
of $\{x_i , X_i^{\pm} \}$, $C(\lambda)$ must be expressed in terms of $\{x_i , X_i^{\pm} \}$.
{}From equations (\ref{Delta1}) and (\ref{kdvdet}) for the quantum determinant,

\be
B(\lambda -h) C(\lambda) = \frac{\lambda}{\lambda-h}D(\lambda-h)A(\lambda) - \lambda^P(\lambda+a)^P. \label{BCdet}
\ee
Consider the first term on the right-hand side;
\be
\frac{\lambda}{\lambda-h}D(\lambda-h)A(\lambda) = \sum_{j=0}^{2P-1} \lambda^{j+1}
\sum_{{k=0} \atop{k \geq j-P \atop{k \leq P-1}}}^{j} \left[ \sum_{l=k}^{P-1} {l \choose k} (-h)^{l-k} D_{l+1}  
\right] A_{j-k},
\ee
where $D_P = 1$ and ${l \choose k}$ denotes the binomial
coefficient, ${l \choose k} = l!/[(l-k)!k!]$.  Whence, substituting $x_n$ for $\lambda$ from the left,

\begin{eqnarray}
\fl C(x_n) = \frac{1}{B_{P-1}} \Bigg(\prod_{l=1}^{P-1} \frac{1}{x_n - x_l -h} \Bigg) \nonumber\\
\times	\Bigg\{ \sum_{j=0}^{2P-1} x_n^{j+1}
\sum_{{k=0} \atop{k \geq j-P \atop{k \leq P-1}}}^{j} \Bigg[ \sum_{l=k}^{P-1} {l \choose k} (-h)^{l-k} D_{l+1} 
 \Bigg] A_{j-k} -x_n^P(x_n+a)^P \Bigg\}. \label{Catx}
\end{eqnarray}
A consideration of the $\lambda^0$ term of $\widetilde{D}(\lambda)$ in equation (\ref{ADup}) reveals that
$A_0 = B_0 w$.  As $w$ is given in terms of the Sklyanin algebra, $\{ x_i,X_i^{\pm} \}$, in equation (\ref{wasxees}),
we have, along with equations (\ref{A}) and (\ref{D}), all of $\{ A_n \}$ and $\{ D_n \}$ in terms of $\{ x_i,X_i^{\pm} \}$.
Therefore equation (\ref{Catx}) gives $C(x_n)$ entirely in terms of the Sklyanin algebra.
Consider now equation (\ref{Bcomup}) with $x_n$ substituted for $\lambda$ from the left, this defines,

	\be
	\widetilde{B}(x_n) = \frac{1}{x_n}\left[(X_n^- - X_n^+)w + \left(1 + \frac{h}{x_n}\right)C(x_n)\right].
		\label{Bx}
	\ee
The right-hand
side of (\ref{Bx}) may be expressed entirely in terms of $\{ x_i,X_i^{\pm} \}$, for brevity denote the right-hand side, 
strictly in terms of $\{ x_i,X_i^{\pm} \}$, by $\widetilde{\mathcal{B}}_n$.
Therefore, with $\mbox{\boldmath $\widetilde{\mathcal{B}}$}_1 = (\widetilde{\mathcal{B}}_1, \widetilde{\mathcal{B}}_2, \ldots,
\widetilde{\mathcal{B}}_{P-1})^t$, we have the following expression,

\be
	\mathbf{\widetilde{B}}_0 = \mathcal{M}^{-1} \mathcal{D} \left( \mbox{\boldmath $\widetilde{\mathcal{B}}$}_1
	-  \mathcal{D}^{P-1} B_{P-1}\mathbf{e}  \right), 
	 \label{xupsym}
\ee
which gives the elementary symmetric polynomials in $\{ \tilde{x}_i \}$ in terms of $\{ x_i,X_i^{\pm} \}$.

Equations (\ref{qddmatrixskly}) (or (\ref{qddmatrix})) and (\ref{xupsym}) constitute the temporal
quantum discrete Dubrovin equations.

	\subsection{Spatial equations} \label{spatialsect}

In this section we consider an evolution along the lattice at a constant 
time level.  Specifically,

	\begin{equation}
	T = L_P L_{P-1} L_{P-2} \ldots L_2 L_1 \mapsto \widehat{T} = L_1 L_P L_{P-1} \ldots L_3 L_2
	\mapsto \cdots  \label{Spatev}
	\end{equation}

\noindent
(The hat, $\widehat{ \,}$ , is used to denote the evolution along the lattice at a constant time-level.)
In the temporal case the reconstruction of $w$ is required for the quantum
discrete Dubrovin equations.  The spatial equations also require the reconstruction of
$1/v_2$,
this is achieved in section \ref{Recspat}.
Together with equation (\ref{wasvees}), this also gives the reconstruction
of $v_1$.  If one requires a reconstruction of the original dynamical variables, $\{ v_{k} \}$, in terms
of the Sklyanin algebra set up at each dynamical variable's particular lattice site (that is, 
$v_1$ and $v_2$ in terms of $\{x_i , X_i^{\pm} \}$,
$v_3$ and $v_4$ in terms of $\{\hat{x}_i , \widehat{X}_i^{\pm} \}$,
etc.)
then the reconstruction is complete.  However, we conjecture that the spatial quantum 
discrete Dubrovin equations, as derived in section \ref{sqdde}, allow for a reconstruction of all of the original dynamical
variables in terms of the unshifted Sklyanin variables $\{x_i , X_i^{\pm} \}$.

		\subsubsection{Reconstruction of $\frac{1}{v_2}$.} \label{Recspat}

In \ref{BCconstant} it is shown that $B_{P-1}$ is equal to the Casimir (\ref{eq:4.2}), 
it is obvious that $\widehat{B}_{P-1}$ also is (as it is still the sum over all $\{ v_{2j} \}$).
Now, from the definitions of the ``conjugate variables'' in section \ref{SV},

\begin{equation}
\fl	\mathbf{A}_{1} - \mathbf{D}_{1} + h\mathbf{D}_{2} \pm2\mathbf{B}_{0}\frac{1}{v_2}
	=   \mathcal{M}^{-1}\left(\mathbf{X}^-_1  + (h{\bf 1}+\mathcal{D})^{-1}
	(h{\bf 1}-\mathcal{D})
	\mathbf{X}^+_1
	- I_0 \mathbf{e} \mp\mathcal{D}^P2B_{P-1}\frac{1}{v_2} \mathbf{e}     \right)  . \label{qdubspat1}
\end{equation}

\noindent
Spatially updated terms follow from (\ref{Spatev}) (or, equivalently, (\ref{Spatup})).
Using the commutation relations in (\ref{TLinv}),
\numparts
	\begin{equation}
\fl	\widehat{A} + \frac{h}{\lambda}\widehat{D} = \frac{1}{\lambda+a}(A - B v_1)(v_2 v_1 + \lambda + a)
	+ \left(1 + \frac{h}{\lambda}\right)\frac{1}{\lambda+a}(C - D v_1)v_2 ,
	\end{equation}

	\begin{equation}
\fl	\widehat{D} = \frac{1}{\lambda+a}[-A v_2 + B (\lambda + a + v_1 v_2)] v_1 
	+ \left(1 + \frac{h}{\lambda}\right)\frac{1}{\lambda+a}[-C v_2 + D (\lambda + a + v_1 v_2)] ,
	\end{equation}

	\begin{eqnarray}
\fl	\widehat{B} = \frac{1}{\lambda(\lambda+a)}[-A v_2 + B (\lambda + a + v_1 v_2)] (v_1 v_2 + h + \lambda + a) \nonumber\\
	+ \left(1 + \frac{h}{\lambda}\right)\frac{1}{\lambda(\lambda+a)}[-C v_2 + D (\lambda + a + v_1 v_2)] v_2 .\label{Bhatlam}
	\end{eqnarray}
\endnumparts
\noindent
So,

	\begin{equation}
\fl	\widehat{A} + \frac{h}{\lambda}\widehat{D} - \widehat{D} + 2 \lambda \widehat{B}\frac{1}{v_2} 
	= -\frac{\lambda + a + 2h}{\lambda+a}A     
	+ \left(1 + \frac{h}{\lambda}\right)D + 2 \frac{(\lambda + a + h)}{\lambda + a} B(\lambda+a+v_1 v_2)\frac{1}{v_2}.\label{spatup}
	\end{equation}

\noindent
{}From equation (\ref{spatup}) one obtains

	\begin{eqnarray}
\fl	\mathbf{\widehat{A}}_{1} - \mathbf{\widehat{D}}_{1} + h\mathbf{\widehat{D}}_{2} + 2\mathbf{\widehat{B}}_{0}\frac{1}{v_2} \nonumber\\
\lo	= \mathcal{M}^{-1}\left(
	\mathbf{X}^+_1 - (\mathcal{D} + a{\bf 1})^{-1}
	(\mathcal{D} + a{\bf 1}+ 2h{\bf 1})\mathbf{X}^-_1
	- I_0 \mathbf{e} -\mathcal{D}^P2 B_{P-1}\frac{1}{v_2}\mathbf{e}
	    \right). \label{qdubspat2}
	\end{eqnarray}

\noindent
It is easily shown
that the leading term in equation (\ref{spatup}) as $\lambda \to \infty$ gives

	\begin{equation}
\fl	2\left( B_{P-1}w + B_{P-1}\frac{h}{v_2} \right) = 
	(\widehat{A}_{P-1} - \widehat{D}_{P-1} + h + 2\widehat{B}_{P-2}\frac{1}{v_2})
	+ (A_{P-1} - D_{P-1} + h -2B_{P-2}\frac{1}{v_2}). 
	\end{equation}
Equations (\ref{qdubspat1}) and (\ref{qdubspat2})
along with (\ref{dub1}), (\ref{dub2}), and (\ref{wAD}) from the reconstruction of $w$, 
and Cramer's rule, lead to

\begin{equation}
\frac{1}{v_2} = \frac{1}{\nu}\left[ 1 + (-1)^{P-1} \sum_{n=1}^{P-1} \frac{1}{x_n} \left( \prod_{{i=1} \atop{i \neq n}}^{P-1} 
	\frac{1}{x_i - x_n}\right)
	\frac{1}{x_n + a}  X_n^-  \right].\label{vxees}
\end{equation}
Along with $w$, as reconstructed in equation (\ref{wasxees}), and equation (\ref{wasvees}) for $w_n$ in
terms of the dynamical variables $\{ v_{k} \}$, equation (\ref{vxees}) leads immediately to
the reconstruction of
$v_1$.

		\subsubsection{Spatial quantum discrete Dubrovin equations.} \label{sqdde}

Equations (\ref{qdubspat1}) and (\ref{qdubspat2}) give

	\begin{eqnarray}
	& 
	\widehat{\mathcal{M}}^{-1}\Big( \mathbf{\widehat{X}^-}_1 + (h{\bf 1}+\widehat{\mathcal{D}})^{-1}
	(h{\bf 1}-\widehat{\mathcal{D}})
	\mathbf{\widehat{X}^+}_1
	- I_0 \mathbf{e} -\widehat{\mathcal{D}}^P2 B_{P-1}\frac{1}{v_2} \mathbf{e} \Big)
	 \,\nonumber\\
	= & \, \mathcal{M}^{-1}\Big( \mathbf{X}^+_1 - (\mathcal{D} + a{\bf 1})^{-1}
	(\mathcal{D} + a{\bf 1}+ 2h{\bf 1})\mathbf{X}^-_1
	- I_0 \mathbf{e} -\mathcal{D}^P2 B_{P-1}\frac{1}{v_2}\mathbf{e} \Big), \label{qddmatrixspat}
	\end{eqnarray}

\noindent
for the spatial part of the quantum discrete Dubrovin equations.
As the invariants of the time evolution are also invariants of the spatial evolution
we may use them, in the form given in
equation (\ref{invcloumn}), to rewrite (\ref{qddmatrixspat}) as

\begin{eqnarray}
	&
	\widehat{\mathcal{M}}^{-1}\Big( (h{\bf 1}+\widehat{\mathcal{D}})^{-1}
	\widehat{\mathcal{D}} \mathbf{\widehat{X}^+}_1
	- \widehat{\mathcal{D}}^P \left(1-B_{P-1}\frac{1}{v_2}\right)\mathbf{e} \Big) 
	 \,\nonumber\\
	= & \,
	\mathcal{M}^{-1}\Big( (\mathcal{D}+a{\bf 1})^{-1}(\mathcal{D}+a{\bf 1}+h{\bf 1})\mathbf{X}^-_1
	- \mathcal{D}^P \left(1-B_{P-1}\frac{1}{v_2}\right)\mathbf{e} \Big). \label{SQDD}
\end{eqnarray}

The expressions for the elementary
symmetric polynomials in the ``spatially updated'' variables, $\{ \hat{x}_i\}$, in terms of
$\{x_i , X_i^{\pm} \}$ are now given. Upon substituting $x_n$ for $\lambda$ from the left, equation (\ref{Bhatlam}) becomes

\begin{eqnarray}
\fl	\widehat{B}(x_n) = \frac{1}{x_n(x_n+a)}X_n^+ w v_2^2 - \frac{1}{x_n(x_n+a)}X_n^- v_2^2 w
	+ \frac{1}{x_n+a}(X_n^+ - X_n^-) v_2 \nonumber\\
	- \left(1 + \frac{h}{x_n} \right)\frac{1}{x_n(x_n+a)}C(x_n)v_2^2.
	\label{Bhatx}
\end{eqnarray}
Note that $C(x_n)$ was expressed in terms of $\{x_i , X_i^{\pm} \}$ by equations (\ref{A}), (\ref{D}),
and (\ref{Catx}); $w$ was reconstructed in terms of $\{x_i , X_i^{\pm} \}$ in equation (\ref{wasxees});
$v_2$ was reconstructed in terms of $\{x_i , X_i^{\pm} \}$ in equation (\ref{vxees}).  Therefore the right-hand side of
(\ref{Bhatx}) can be given entirely in terms of $\{x_i , X_i^{\pm} \}$.  For brevity denote the
right-hand side of (\ref{Bhatx}) expressed strictly in terms of $\{x_i , X_i^{\pm} \}$
by $\widehat{\mathcal{B}}_n$.  With this notation

\be
	\mathbf{\widehat{B}}_0 = \mathcal{M}^{-1} \mathcal{D} \left( \mbox{\boldmath $\widehat{\mathcal{B}}$}_1
	-  \mathcal{D}^{P-1} B_{P-1}\mathbf{e}  \right), 
	 \label{xspaceupsym}
\ee
giving the elementary symmetric polynomials in $\{ \hat{x}_i\}$ in terms of $\{ x_i, X_i^{\pm}\}$.

Equations (\ref{SQDD}) (or (\ref{qddmatrixspat})) and (\ref{xspaceupsym}) constitute the spatial quantum discrete
Dubrovin equations.

		\section{Examples}\label{exs}

The well-defined temporal evolution which follows from the quantum
discrete Dubrovin equations is illustrated
for the $P=2$ case in section \ref{p2ex} and for the $P=3$ case in section \ref{p3ex}.		
The reconstruction of the original dynamical variables, $\{ v_{k} \}$,
in terms of the Sklyanin algebra variables, $\{ x_i, X_i^{\pm} \}$,
is also performed in the $P=2$ and $P=3$ cases.
The reconstruction is achieved, essentially, via a ``spatial evolution'' using 
the spatial part of the quantum discrete Dubrovin equations.

	\subsection{The $P = 2$ case}\label{p2ex}
	
		\subsubsection{Temporal evolution}
		
With $P=2$, equation (\ref{wasxees}) gives

	\be
	w = \frac{1}{\nu}\left( \nu^2 + 2a + h + 2x -\frac{1}{x+h}X^+ - \frac{1}{x}X^-   \right).
	\label{wP2}
	\ee
Equation (\ref{qddmatrixskly}) gives directly that

	\be
	\tilde{x} -\frac{1}{\tilde{x}+h}\widetilde{X}^+ = x - \frac{1}{x}X^-.
	\label{qddP2}
	\ee
Along with (\ref{wP2}), equation (\ref{xupsym}) leads to

	\be
	\tilde{x} = \frac{1}{\nu} w \left( x - \frac{1}{x}X^- \right).
	\label{symP3}
	\ee
It is easily seen that 

	\begin{displaymath}
	\left( 1 + \frac{h}{x}  \right)\frac{1}{x+h}X^+ = -\frac{1}{x}X^- + \frac{1}{x}I_0 + I_1 +2x.
	\end{displaymath}
Hence we may consider the evolution given by the quantum discrete Dubrovin equations to be that
of $x$ and $\frac{1}{x+h}X^+$ (or $\frac{1}{x}X^-$) along with the preservation of the invariant,
$I_0 = \widetilde{I}_0 \ldots$.  Equation (\ref{IPminus1}) gives $I_1 = \nu^2 + 2(a+h)$.

		\subsubsection{Reconstruction}
	
Setting $P=2$ in equation (\ref{vxees}) gives

	\be
	\frac{1}{v_2} = \frac{1}{\nu}\left( 1 - \frac{1}{x(x+a)}X^-  \right).
	\label{vP2}
	\ee
It follows trivially from the Casimir (\ref{eq:4.2}), $v_2 + v_4 = \nu$, that
we also have $v_4$ in terms of $(x, X^{\pm})$.
Equation (\ref{wasvees}) along with (\ref{wP2}) and (\ref{vP2}) gives the reconstruction of $v_1$,
and the Casimir (\ref{eq:4.2}), $v_1 + v_3 = \nu$, gives the reconstruction of $v_3$.
Therefore, full reconstruction has been achieved.

	\subsection{The $P = 3$ case}\label{p3ex}

		\subsubsection{Temporal evolution}
		
For this section we define

	\begin{displaymath}
\fl	\pi_1  := \frac{1}{x_2-x_1}\left(\frac{x_2}{x_1+h}X^+_1 - \frac{x_1}{x_2+h}X^+_2  \right) 
	\qquad \pi_2  := \frac{1}{x_2-x_1}\left(\frac{1}{x_1+h}X^+_1 - \frac{1}{x_2+h}X^+_2  \right),
	\end{displaymath}

	\begin{displaymath}
\fl	\phi_1  := \frac{1}{x_2-x_1}\left(\frac{x_2}{x_1}X^-_1 - \frac{x_1}{x_2}X^-_2  \right) 
	\qquad \phi_2  := \frac{1}{x_2-x_1}\left(\frac{1}{x_1}X^-_1 - \frac{1}{x_2}X^-_2  \right).
	\end{displaymath}

\noindent
Then, with $P=3$, equation (\ref{wasxees}) gives

	\be
	w = \frac{1}{\nu}\left[ \nu^2 + 3a + 2h + 2(x_1+x_2) + \pi_2 + \phi_2  \right].
	\label{wP3}
	\ee
Equation (\ref{qddmatrixskly}) gives directly that

	\be
	\tilde{x}_1\tilde{x}_2 + \tilde{\pi}_1  = x_1x_2 + \phi_1,
	\label{qdd1P3}
	\ee

	\be
	\tilde{x}_1 + \tilde{x}_2 + \tilde{\pi}_2  = x_1 + x_2 + \phi_2.
	\label{qdd2P3}
	\ee
Along with (\ref{wP3}), equation (\ref{xupsym}) leads to

	\be
	\tilde{x}_1\tilde{x}_2 = \frac{1}{\nu} w \left( x_1x_2 + \phi_1 \right),
	\label{sym1P3}
	\ee

\begin{eqnarray}
	&\tilde{x}_1 + \tilde{x}_2 =  & \frac{1}{\nu} w \left( x_1 + x_2 + \phi_2 \right)
	- \frac{1}{\nu^2}[ 2x_1x_2 +h(x_1+x_2) + h^2 - 3a^2
	 \,\nonumber\\
	& & -(x_1+x_2+2h)(x_1+x_2+2h+3a) + \pi_1 + \phi_1 + \pi_2\phi_2 ].\label{sym2P3}
\end{eqnarray}
It is easily seen that 

	\begin{displaymath}
	\frac{h}{x_1x_2}\pi_1 + \pi_2 = -\phi_2 + \frac{1}{x_1x_2}I_0 - I_2 -2(x_1+x_2)
	\end{displaymath}
and

	\begin{displaymath}
	\left[1+\frac{h(x_1+x_2)}{x_1x_2}\right]\pi_1 - h\pi_2 = -\phi_1 + \frac{x_1+x_2}{x_1x_2}I_0 + I_1 -2x_1x_2.
	\end{displaymath}	
Hence we may consider the evolution given by the quantum discrete Dubrovin equations to be that
of the elementary symmetric polynomials $x_1x_2$ and $x_1 + x_2$, and $\pi_1$ and $\pi_2$ (or $\phi_1$ and $\phi_2$), 
along with the preservation of the invariants,
$I_i = \widetilde{I}_i \ldots$.  The form of the invariants follows from (\ref{IPminus1}) and (\ref{taunoughtfmla}).

		\subsubsection{Reconstruction}

Setting $P=3$ in equation (\ref{vxees}) gives

	\begin{equation}\label{v2p3}
	\frac{1}{v_2} = \frac{1}{\nu}
	\left[1+ \frac{1}{x_2-x_1}\left( \frac{1}{x_1(x_1+a)}X_1^- - \frac{1}{x_2(x_2+a)}X_2^-   \right)   \right].
	\end{equation}
In the $P=3$ case, equation (\ref{SQDD}) of the spatial part of the quantum discrete Dubrovin equations reads

\begin{eqnarray}\label{qdds2}
	& 
	\left( \begin{array}{c} 
	-\hat{x}_1\hat{x}_2\left(1-\frac{\nu}{v_2}\right) - \frac{1}{\hat{x}_2-\hat{x}_1}
	\left( \frac{\hat{x}_2}{\hat{x}_1+h}\widehat{X}_1^+ - \frac{\hat{x}_1}{\hat{x}_2+h}\widehat{X}_2^+    \right)\\ 
	(\hat{x}_1+\hat{x}_2)\left(1-\frac{\nu}{v_2}\right) + \frac{1}{\hat{x}_2-\hat{x}_1}
	\left( \frac{1}{\hat{x}_1+h}\widehat{X}_1^+ - \frac{1}{\hat{x}_2+h}\widehat{X}_2^+ \right) \end{array}
	\right) \,\nonumber\\
	= & \,  \left( \begin{array}{c} 
	-x_1x_2\left(1-\frac{\nu}{v_2}\right) - \frac{1}{x_2-x_1}
	\left[ \frac{x_2}{x_1} \left( 1 + \frac{h}{x_1+a} \right) X_1^- - \frac{x_1}{x_2}\left( 1 + \frac{h}{x_2+a} \right)X_2^-  \right] \\ 
	(x_1+x_2-h)\left(1-\frac{\nu}{v_2}\right) + \frac{1}{x_2-x_1}
	\left( \frac{1}{x_1} X_1^- - \frac{1}{x_2}X_2^-    \right) \end{array} \right). 
\end{eqnarray}
The top $-(a+h)$bottom of (\ref{qdds2}) leads to

	\begin{eqnarray}
\fl	[-\hat{x}_1\hat{x}_2-(a+h)(\hat{x}_1+\hat{x}_2)]\left(1-\frac{\nu}{v_2}\right) - \frac{1}{\hat{x}_2-\hat{x}_1}
	\left( \frac{\hat{x}_2+a+h}{\hat{x}_1+h}\widehat{X}_1^+ - \frac{\hat{x}_1+a+h}{\hat{x}_2+h}\widehat{X}_2^+    \right)\nonumber\\
\lo	= (a+h)^2\left(1 - \frac{\nu}{v_2}\right).
	\end{eqnarray}
Therefore,

	\begin{displaymath}
	\frac{1}{v_0} = \frac{1}{\nu}\left[ 1+ \frac{1}{x_2-x_1}\left( \frac{1}{(x_1+h)(x_1+a+h)}X_1^+
	- \frac{1}{(x_2+h)(x_2+a+h)}X_2^+  \right)   \right].
	\end{displaymath}
It follows trivially from the Casimir (\ref{eq:4.2}),
$v_0 + v_2 + v_4 = \nu$,
that we also have $v_4$ in terms of $\{ x_i , X_i^{\pm} \}_{i=1,2}$.

The reconstruction of $\{ v_{2j+1} \}$ follows from that of $w$.
It follows from equation (\ref{wasvees}) that $v_1$, in terms of $\{ x_i , X_i^{\pm} \}_{i=1,2}$,
is obtained from equations (\ref{wP3}) and (\ref{v2p3}).
Using the form of $I_0$ which follows from (\ref{taunoughtfmla}), one obtains,

	\begin{displaymath}
	w = \frac{1}{\nu} \left[ \frac{1}{x_1x_2}I_0
	-h -\frac{h}{x_1x_2}\pi_1    \right].
	\end{displaymath}
Now, as $I_0$ is invariant under spatial updates, it may be written, using the same expression from equation
(\ref{taunoughtfmla}), at any spatial level.  In simple terms, as $I_0 = \widehat{I}_0 = \widehat{\widehat{I}}_0 = \ldots$,
the same expression for $I_0$ may be used with any number of hats above or below the operators.  Hence,

	\begin{equation}\label{wupP3}
	\hat{w} = \frac{1}{\nu} \left[ \frac{1}{\hat{x}_1\hat{x}_2}I_0
	-h - \frac{h}{\hat{x}_1\hat{x}_2}\hat{\pi}_1    \right].
	\end{equation}
The top equation of the spatial quantum discrete Dubrovin equations, (\ref{qdds2}), gives

	\begin{eqnarray}
\fl	\frac{1}{\hat{x}_1\hat{x}_2}\hat{\pi}_1
	=  (\frac{1}{\hat{x}_1\hat{x}_2}x_1x_2-1)\left(1 - \frac{\nu}{v_2}\right) \nonumber\\
	+\frac{1}{\hat{x}_1\hat{x}_2}\frac{1}{x_2-x_1}\left[ \frac{x_2}{x_1}\left(1+\frac{h}{x_1+a}\right)X_1^-
	- \frac{x_1}{x_2}\left(1+\frac{h}{x_2+a}\right)X_2^-  \right], 
	\end{eqnarray}
and equation (\ref{xspaceupsym}) gives,

\begin{eqnarray}\label{xupprodP3}
	&\hat{x}_1\hat{x}_2 =  & - \frac{1}{a \nu}\left(x_1 x_2 + \phi_1 \right) v_2^2w + \frac{1}{a}x_1 x_2 v_2 w
	+ \frac{h}{a \nu}\left(x_1 x_2 + \pi_1 \right) v_2
	 \,\nonumber\\
	& & - \frac{h}{a \nu^2}[ 2x_1x_2 +h(x_1+x_2) + h^2 - 3a^2 
	 \,\nonumber\\
	& & -(x_1+x_2+2h)(x_1+x_2+2h+3a) + \pi_1 + \phi_1 + \pi_2\phi_2 ]v_2^2.
\end{eqnarray}
So, with equations (\ref{wP3}) and (\ref{v2p3}), this gives $\hat{x}_1\hat{x}_2$ in terms of $\{x_i , X_i^{\pm} \}$.
Therefore we have reconstructed $\hat{w}$ in terms of $\{ x_i, X_i^{\pm} \}$.
With the reconstruction of $v_4$ this then gives a reconstruction of $v_3$, and from the Casimir (\ref{eq:4.2}),
$v_1 + v_3 + v_5 = \nu$,
this gives a reconstruction of $v_5$.  Therefore, full reconstruction has been achieved.

		\section{Conclusion}

The quantum discrete Dubrovin equations have been derived.  Equations
(\ref{qddmatrixskly}) and (\ref{xupsym}) give a temporal evolution and
(\ref{SQDD}) and (\ref{xspaceupsym}) give a spatial evolution.

The classical discrete Dubrovin equations were published in \cite{Ni:discdub} and
\cite{NiEn:Intmaphyp}, but in \cite{Ni:discdub} the focus was on temporal equations only.
In the temporal case, the work presented here is very much analogous to the classical case.
The classical
($h \to 0$) limit of (\ref{qddmatrix}) is the discrete Dubrovin equations as presented in \cite{Ni:discdub}.  This
is seen as follows.  
The derivation of the (classical) discrete Dubrovin equations employs
the invariant spectral curve
\be
		\textrm{det}(T(\lambda) - \eta) = 0, \label{character}
\ee
which defines a hyperelliptic curve of genus $g = P-1$. 
The classical equations
are written in terms of the discriminant of the hyperelliptic
curve (\ref{character}), $R(\lambda)$.  This discriminant may be expressed as

	\be
	R(\lambda) = \frac{1}{4}( A(\lambda) + D(\lambda) )^2 - \textrm{det}(T(\lambda)).\label{classR}
	\ee
Following \cite{Ni:discdub}, we see that (\ref{classR}) implies that
\begin{displaymath}	
	\frac{1}{2}( A(\lambda) - D(\lambda) ) = \kappa \sqrt{R(\lambda) - B(\lambda)C(\lambda)},
\end{displaymath}
where $\kappa$ denotes the sign $\pm$ and corresponds to the choice of sheet of the Riemann surface,
the condition $\tilde{\kappa} = - \kappa$ can be seen by also considering (\ref{UPDATE}) and 
evaluating at the (no longer \emph{operator}) roots of $B(\lambda)$.
A quantum deformation of the expression (\ref{classR}), evaluated at an operator root of $B(\lambda)$, $x_n$, is

	\begin{equation}
R(x_n) = \frac{1}{4} \tau(x_n)^2 - \frac{1}{2}\left(\Delta(x_n - \frac{h}{2}) + \Delta(x_n + \frac{h}{2})     \right) 
= \frac{1}{4}\left( X_n^- - X_n^+ \right)^2.
	\end{equation}

\noindent
Therefore, in the classical limit, the quantum discrete Dubrovin equations, (\ref{qddmatrix}), become

\begin{equation}
	\widetilde{\mathcal{M}}^{-1}\Big( \tilde{\kappa}  \sqrt{\mathbf{R}(\tilde{\mathbf{x}}_1)} 
	- \frac{1}{2}I_0 \mathbf{e}  \Big) = 
	\mathcal{M}^{-1}\Big( - \kappa \sqrt{\mathbf{R}(\mathbf{x}_1)} 
	- \frac{1}{2}I_0 \mathbf{e}  \Big). \label{cddmatrix}
\end{equation}

\noindent
where $\sqrt{\mathbf{R}(\mathbf{x}_1)} = (\sqrt{R(x_1)}, \ldots, \sqrt{R(x_{P-1})})^t$. Equation (\ref{cddmatrix})
is the classical discrete Dubrovin equation, as first presented in \cite{Ni:discdub}.
(The notation of \cite{Ni:discdub} is such that the roots of $B(\lambda)$ are denoted by $\{\mu_i \}$, rather than
by $\{x_i \}$.)
Classically, the discrete Dubrovin equations are the intermediate step towards
the parameterization of the orbits of the classical map.
In \cite{Ni:discdub} the parametrization of the solutions of (\ref{cddmatrix}) in terms of Abelian
functions of Kleinian type was discussed, and illustrated in the $P=2$ and $P=3$ cases.
{}From the new perspective of this paper it is surprising that the classical limit of (\ref{xupsym})
is not required for this parametrization, and, indeed, does not feature in \cite{Ni:discdub}.

Returning to the quantum equations, the next issues to be addressed concern the representation
theory.  Then, drawing inspiration from the classical discrete Dubrovin equations, we
would hope to be able to construct explicit expressions for the quantum
propagators interpolating over an arbitrary number of discrete-time steps.  Effective mechanisms for computing expectation
values and long-time asymptotics for the transition amplitudes would also be 
desirable corollaries of this proposed work.  
Extensions of the present work to the higher rank case associated
with quantum mappings in the Gel'fand-Dikii hierarcy \cite{NiCa:iqm}, \cite{NiCa:fusion}, 
also form the subject of future work.

		\ack

CMF would like to thank the EPSRC for financial support.
FWN is thankful to V Enolskii for discussions on the classical
aspects of this work.

	\appendix

	\section{Quantum invariants}\label{invproof}
	
Consider the following evolution of the monodromy matrix: 

	\begin{equation}
	T' = N T N^{-1}      \label{Nevolve}
	\end{equation}

\noindent
(so, to be specific, $N$ could be $M_1$ or $L_1$).
In this appendix it will be proven that if $N$ satisfies the relations

\be
R^+_{12}\,N_1 N_2 = N_2 N_1 R^-_{12} \   \label{RN}
\ee
and 

\be
T_1 \, N_1^{-1} \,S^{+}_{12}\, N_2 \ =\ N_2 \, S^{-}_{12}\,
T_1\, N_1^{-1}\     \label{TNinv}
\ee

\noindent
then the Yang-Baxter relation (\ref{RTST}) is preserved through this
evolution and, moreover,

	\begin{displaymath}
	\tau( \lambda )\ =\ \tr\left( K(\lambda )T(\lambda )\right)\  
	\end{displaymath}

\noindent
is invariant under this evolution, for a certain $K(\lambda)$ which will
also be derived.
First the preservation of the Yang-Baxter relation (\ref{RTST}) under 
the evolution (\ref{Nevolve}) is shown.
It requires only (\ref{RN}), (\ref{TNinv}), and the constraints on the $R$ and $S$ matrices that
$R_{12}^{\pm}S_{12}^{\pm} = S_{12}^{\mp} R_{12}^{\mp}$,

\bea  
R^+_{12} T^{\prime}_1 \, S^+_{12} T^{\prime}_2 &=& 
R^+_{12} N_1\, T_1\, N_1^{-1}\, S^+_{12} N_2\, 
T_2\, N_2^{-1}  \nn \\
&=& R^+_{12} N_1\, N_2\, S^-_{12} T_1\, N_1^{-1}\, T_2\, 
N_2^{-1}  \nn \\
&=& N_2\, N_1\, S^+_{12}R^+_{12} T_1\, N_1^{-1}\, T_2\, 
N_2^{-1}  \nn \\
&=& N_2\, N_1\, S^+_{12}R^+_{12} T_1 \, S^+_{12} T_2 \, 
N_2^{-1}\, N_1^{-1} S_{12}^{-^{-1}}  \nn \\
&=& N_2\, N_1\, S^-_{21} T_2\, S^-_{12} T_1\, R^-_{12} 
N_2^{-1}\, N_1^{-1} S_{12}^{-^{-1}}  \nn \\
&=& T^{\prime}_2\, S_{21}^+ N_1\, N_2\, S_{12}^- T_1 
N_1^{-1}\, N_2^{-1} S_{12}^{+^{-1}} R^-_{12}  \nn \\
&=& T_2^{\prime}\, S_{12}^- T_1^{\prime} R^-_{12}  \  .  
\eea

\noindent
The derivation of the invariants of the evolution (\ref{Nevolve}) will now be given.
Introduce the tensor
\be 
K_{12}\ =\ P_{12}K_1K_2\   ,  
\ee 
where $P_{12}$ is the permutation operator, which satisfies the relations
\be
P_{12} \, ( A \otimes B)  =  (B \otimes A) \, P_{12} \qquad 
P_{12} = 
P_{21} \qquad \tr_2 P_{12} = {\bf 1}_1.
\ee
Choosing 
$\lambda_1 = \lambda_2$, we can take the trace
of (\ref{TNinv}) contracted with $K_{12}$. The left-hand side leads to 
\be
\tr_{1} \tr_{2} \left(K_{12} T_1 N^{-1}_1 \, S_{12}^+ N_2\right) \   =\ \tr_{2}\left( K_2 T_2 
N_2^{-1}\,\tr_1(P_{12}K_2S_{12}^+)\,N_2\right)\ =\ \tr(KT) 
\ee
provided that 
\be 
\tr_1(P_{12}K_2S_{12}^+)={\bf 1}_2 \  . \label{PKS}
\ee 
Under the same condition, 
(\ref{PKS}), we have, from the right-hand side of equation (\ref{TNinv}), that
\be 
\tr_{1} \tr_2 \left( K_{12}N_2 \, S_{12}^- T_1 N^{-1}_1\right)  
 =\ \tr_{1}\left( K_1 
N_1\,\tr_2(P_{12}K_1S_{12}^-)\,T_1 N^{-1}_1\right)\ =\ 
\tr(KT^{\prime})\  .  
\ee 
Hence, if $K(\lambda)$ is a solution of equation (\ref{PKS}), we have the invariance 
of $\tr(K(\lambda)T(\lambda))$ under the evolution
given by equation (\ref{Nevolve}).

A solution to equation (\ref{PKS}) is found 
by taking
\be \label{eq:PS} 
K_2\,=\,\tr_1\left\{ P_{12} \,^{t_1\!}\left[ ( \,^{t_1\!}S^+_{12})^{-1}
\right] \right\} \    . \label{PSsoln}
\ee 
This is most easily verified by introducing the twisted product

	\begin{displaymath}
	X_{1 2} * Y_{1 2} := ^{t_2\!}(^{t_2\!}X_{1 2} \, ^{t_2\!} Y_{1 2})
	= ^{t_1\!}(^{t_1\!}Y_{1 2} \, ^{t_1\!}X_{1 2}).
	\end{displaymath}
An inverse  with respect to the $*$ product is

	\begin{displaymath}
	X_{1 2}^{-1 *} = ^{t_2\!}((^{t_2\!}X_{1 2})^{-1}) = ^{t_1\!}((^{t_1\!}X_{1 2})^{-1}).
	\end{displaymath}
So, with equation (\ref{PSsoln}),

	\begin{eqnarray}
	\tr_1(P_{12}K_2S_{12}^+) 
	& = & \tr_1 \tr_3 (P_{12} P_{32} (S_{32}^+)^{-1*} S_{12}^+) \nonumber\\
	& = & \tr_1 \tr_3 (P_{32} P_{13} (S_{32}^+)^{-1*} S_{12}^+) \nonumber\\
	& = & \tr_3 (P_{32} \, \tr_1(P_{13} (S_{32}^+)^{-1*} S_{12}^+)) \nonumber\\
	& = & \tr_3 (P_{32} \, S_{32}^+ * (S_{32}^+)^{-1*}) \nonumber\\
	& = & \tr_3 (P_{32} \, {\bf 1}_{3 2}) = {\bf 1}_2.\nonumber
	\end{eqnarray}

\noindent
For the proof of the commutativity of the invariants $\tau(\lambda) = \tr(K(\lambda)T(\lambda))$
with the form of $K(\lambda)$ given in equation (\ref{PSsoln}) we refer the reader to \cite{NiCa:Quantof}.

	\section{Quantum determinant factorization}\label{qdetappan}

In the case of ultralocal models it is straightforward to show that the quantum determinant
is equal to the product of the local quantum determinants of the
constituent $L$ operators.  This factorization will now be proven for the present
non-ultralocal case, which is less straightforward.

Equations (\ref{eq:RLL}) to (\ref{eq:2.1c}) and (\ref{RSSR}) give

\be
R_{12}^+\,\stackrel{\longleftarrow}{\prod_{j=1}^{m}}\ 
L_{j,1}\ \,\stackrel{\longleftarrow}{\prod_{k=1}^{m}}\ 
L_{k,2}\ = 
\stackrel{\longleftarrow}{\prod_{j=1}^{m}}\ 
L_{j,2} \, \stackrel{\longleftarrow}{\prod_{k=1}^{m}}\ 
L_{k,1}\ \,R_{12}^-\  , \label{eq:RpiLpiL}  
\ee
for $m \le P - 1$. 
{}From equations (\ref{eq:RpiLpiL}) and (\ref{RSSR}),

\be
R_{12}^- S_{12}^- \,\stackrel{\longleftarrow}{\prod_{j=1}^{m}}\ 
L_{j,1}\ \,\stackrel{\longleftarrow}{\prod_{k=1}^{m}}\ 
L_{k,2}\ = 
S_{12}^+ \stackrel{\longleftarrow}{\prod_{j=1}^{m}}\ 
L_{j,2} \, \stackrel{\longleftarrow}{\prod_{k=1}^{m}}\ 
L_{k,1}\ \,R_{12}^-  . \label{qdetRpiLpiL}  
\ee
Therefore, for the particular relative value of the spectral parameters $\lambda_1$ and $\lambda_2$
such that $R^-_{12}$ is proportional to a rank-one projector, equation (\ref{qdetRpiLpiL}) is defined to be
equal to 
$R_{12}^- \textrm{Qet}(\stackrel{\longleftarrow}{\prod_{j=1}^{m}}\ L_{j} )$.  
Consider the left-hand side of equation (\ref{qdetRpiLpiL}),

	\begin{eqnarray}
	R_{12}^- \textrm{Qet}(\stackrel{\longleftarrow}{\prod_{j=1}^{m}}\ L_{j} ) 
	& = & R_{12}^- S_{12}^- \, \stackrel{\longleftarrow}{\prod_{j=1}^{m}}\ 
	L_{j,1}\ \,\stackrel{\longleftarrow}{\prod_{k=1}^{m}}\ 
	L_{k,2}\ \nonumber\\
	& = & R_{12}^- S_{12}^- \, L_{m,1}\,  L_{m,2}\, S_{21}^+  \stackrel{\longleftarrow}{\prod_{j=1}^{m-1}}\ 
	L_{j,1}\ \, \stackrel{\longleftarrow}{\prod_{k=1}^{m-1}}\ 
	L_{k,2}\ \nonumber\\
	& = & S_{12}^+  \, L_{m,1}\,  L_{m,2}\, R_{12}^- S_{12}^-  \stackrel{\longleftarrow}{\prod_{j=1}^{m-1}}\ 
	L_{j,1}\ \, \stackrel{\longleftarrow}{\prod_{k=1}^{m-1}}\ 
	L_{k,2}\ \nonumber\\
	& = & \textrm{Qet}(L_{m}) R_{12}^- \textrm{Qet}(\stackrel{\longleftarrow}{\prod_{j=1}^{m-1}}\ 
	L_{j,1}),
	\end{eqnarray}

\noindent
and, therefore,

\be
\textrm{Qet} \left(\stackrel{\longleftarrow}{\prod_{j=1}^{m}}\ 
L_j \right) = \stackrel{\longleftarrow}{\prod_{j=1}^{m}}\ 
\textrm{Qet}(L_j),   \label{detproduct}
\ee
where $m \le P -1$. The proof of the same result via the right hand side of (\ref{qdetRpiLpiL}) follows
similarly.

The quantum determinant for the Yang-Baxter structure (\ref{RTST}),
$\Delta$, was introduced in equation
(\ref{RTSTdet}).
{}From the left-hand side of (\ref{SRTST}),

	\begin{eqnarray}
	R_{12}^- \Delta& = & R_{12}^- S_{12}^-  \,T_1 \,S^{+}_{12}\, T_2 \nonumber\\
	& = & R_{12}^- S_{12}^-  \,L_{P,1} \, L_{P,2} \,S^{+}_{21} \, \stackrel{\longleftarrow}{\prod_{j=1}^{P-1}}\ 
L_{j,1} \, \stackrel{\longleftarrow}{\prod_{k=1}^{P-1}}\ 
L_{k,2}          \nonumber\\
	& = & S_{12}^+  \,L_{P,2} \, L_{P,1} R_{12}^-  \, S^{-}_{12} \, \stackrel{\longleftarrow}{\prod_{j=1}^{P-1}}\ 
L_{j,1} \, \stackrel{\longleftarrow}{\prod_{k=1}^{P-1}}\ 
L_{k,2}  ,  \label{RTSTqdet} 
	\end{eqnarray}
(the same result follows similarly for the right-hand side) and hence, from equation (\ref{detproduct}),
we have

\begin{displaymath}
\Delta = \stackrel{\longleftarrow}{\prod_{j=1}^{P}}\ 
\textrm{Qet}(L_j).  
\end{displaymath}

	\section{$B_{P-1}$ and $C_{P}$ are in the centre of the algebra}\label{BCconstant}

It can be shown, using the commutation relations (\ref{RTST}) with the explicit realization of the $R$ and $S$
matrices (\ref{eq:Rsol}) and the gradation of the monodromy matrix (\ref{Tgraded}),
that $B_{P-1}$ and $C_P$ belong to the centre of the algebra.
The model specific proof given here, however, is more constructive for our
purposes as it reveals $B_{P-1}$ and $C_P$ to have the value prescribed to the Casimirs in equation (\ref{eq:4.2}).
Consider $L_{P+1}$,

\be
L_{P+1}\ =\ \left( \begin{array}{cc} \lambda + a + v_{2P+2} v_{2P+1}
& v_{2P+2} \\ 
\lambda v_{2P+1} 
& \lambda \end{array} \right) \ . \label{LPplus1}
\ee
{}From the definition of the monodromy matrix, equation (\ref{Tmatrix}), and the grading of the monodromy
matrix for period $P$, equation (\ref{Tgraded}), along with equation (\ref{LPplus1}), we observe that the 
monodromy matrix of period $P+1$,
$T_{P+1}$, has the form
\begin{displaymath}
\fl	\left( \begin{array}{cc} \lambda^{P+1} + \lambda^{P} [A_{P-1} + a + v_{2P+2} v_{2P+1}
+ v_{2P+2}C_P] + \ldots 
& \lambda^{P} [B_{P-1}+ v_{2P+2}] + \ldots \\ 
\lambda^{P+1} [C_{P} + v_{2P+1}] + \ldots 
& \lambda^{P+1} + \lambda^{P} [D_{P-1} + v_{2P+1}B_{P-1}] + \ldots \end{array} \right) \ . \label{TPplus1}
\end{displaymath}

\noindent
The $B$ and $C$ entries give us recursion relations for the coefficients of the highest order of $\lambda$. 
Using these, in conjunction with the information from (\ref{LPplus1}) that for the 
period $=1$ case $C_P = v_1$ and $B_{P-1} = v_2$, we obtain
that for the period $=P$ case

\be
B_{P-1} \ =\ \sum_{n=1}^{P} v_{2n} \qquad
C_{P} \ =\ \sum_{n=1}^{P} v_{2n-1}. \label{BandC}
\ee
For the mappings of KdV type these summations 
are the Casimirs given in equation (\ref{eq:4.2}).

\end{document}